\documentclass[prc,aps,twocolumn,preprintnumbers,amsmath,amssymb,floatfix,superscriptaddress]{revtex4}
\usepackage[usenames]{color}
\usepackage{amsmath}
\usepackage{graphicx}
\usepackage{float,epsfig}
\usepackage{lscape}
\newcommand\be{\begin{equation}}
\newcommand\ee{\end{equation}}
\newcommand\bea{\begin{eqnarray}}
\newcommand\eea{\end{eqnarray}}

\usepackage[usenames]{color}
\usepackage{ulem}

\begin{document}
\title{Nuclear excitations as coupled one and two random--phase--approximation modes}
\author{D. Gambacurta}
\affiliation{Extreme Light Infrastructure - Nuclear Physics (ELI-NP), “Horia Hulubei”
 National Institute for Physics and Nuclear Engineering, 30 Reactorului
 Street, RO-077125 Magurele, Jud. Ilfov, Romania}
\affiliation{I.N.F.N. - Sezione di Catania, 95123 Catania, Italy}

\author{F. Catara}
\affiliation{Dipartimento di Fisica e Astronomia, Universit\'a di Catania, Italy}
\affiliation {I.N.F.N. - Sezione di Catania, 95123 Catania, Italy}

\author{M. Grasso}
\affiliation{Institut de Physique Nucl\'eaire, IN2P3-CNRS, Universit\'e Paris-Sud, 
F-91406 Orsay Cedex, France}

\author{M. Sambataro}
\affiliation {I.N.F.N. - Sezione di Catania, 95123 Catania, Italy}

\author{M. V. Andr\'es}
\affiliation{Departamento de F\'{i}sica Atomica, Molecular y Nuclear, 41080 Sevilla, Spain}

\author{E. G. Lanza}
\affiliation {I.N.F.N. - Sezione di Catania, 95123 Catania, Italy}
\affiliation{Dipartimento di Fisica e Astronomia, Universit\'a di Catania, Italy}

\begin{abstract}

We present an extension of the random--phase approximation (RPA) where the RPA 
phonons are used as building blocks to construct the excited states. In our 
model, that we call double RPA (DRPA), we include up to two RPA phonons. This is 
an approximate and simplified way, with respect to the full second random--phase 
approximation (SRPA), to extend the RPA by including two particle--two hole 
configurations. Some limitations of the standard SRPA model, related to the 
violation of the stability condition, are not encountered in the DRPA. We also 
verify in this work that the energy--weighted sum rules are satisfied. The DRPA is applied to low--energy modes and giant resonances in the nucleus 
$^{16}$O. We show that the model (i) produces a global downwards shift of the 
energies with respect to the RPA spectra; (ii) provides a shift that is however 
strongly reduced compared to that generated by the standard SRPA. This model 
represents an alternative way of correcting for the SRPA anomalous energy shift, 
compared to a recently developed extension of the SRPA, where a subtraction 
procedure is applied. The DRPA provides results in good agreeement with the experimental energies, 
with the exception of those low--lying states that have a dominant two 
particle--two hole nature. For describing such states, higher--order 
calculations are needed. 
\end{abstract}

\maketitle

\section{Introduction}
  A common feature of many-body systems is the presence of collective modes in 
their excitation spectra. Particularly interesting are those which can be 
interpreted in terms of vibrations. In nuclei, they have been known for a long 
time\;\cite{BM} and are observed both in the low--lying and in the 
higher--energy region (giant resonances (GR))\;\cite{BM,HV}. Another example of 
collective vibrations is the dipole plasmon resonance in metallic 
clusters\;\cite{dH,Br}, a very collective mode which is the analogue of the 
nuclear dipole GR and is due to the vibration of the center of mass of the 
delocalized valence electrons against that of the positive ions.

 In all these cases, several characteristic properties are rather well 
reproduced within the random--phase approximation (RPA), which is for this 
reason considered a good microscopic theory for the study of collective states 
in many-body systems\;\cite{Ro,RS}. Among its merits, we mention that the RPA 
preserves the energy--weighted sum rules (EWSR), in the sense of the Thouless 
theorem\;\cite{Th}. This feature is very important because it guarantees that 
spurious states associated with broken symmetries are exactly separated in RPA 
from the physical states of the system.

 On the other hand, the RPA model has some limits. Among them we recall that, by 
construction, it predicts a perfectly harmonic spectrum and cannot reproduce the 
spreading width of the excited modes. As underlined in several studies, the 
coupling of one particle--one hole (1p-1h) configurations with more complex 
states is very important to overcome these limitations and to describe the 
spreading widths and the anharmonicities in the excitation spectra (see, for 
example Refs.\;\cite{BCB,Li,La1} and references therein). A natural extension of 
the RPA is the so--called second RPA (SRPA), which amounts to enlarge the space 
of basic elementary excitations by 
 including two particle--two hole (2p-2h) configurations and by coupling them 
with the 1p-1h ones and among themselves. It was shown\;\cite{Ya,AL,Pa}  that 
the EWSR are exactly the same in SRPA and in RPA. 

Recent applications of the SRPA model were performed without any cut and 
approximations in the matrices. Such applications provided a very strong 
modification (a large downwards energy shift) of the RPA excitation spectra, 
also for those collective states 
whose description within the RPA is rather good. This was obtained 
for collective nuclear excitations with  Skyrme and Gogny 
forces\;\cite{GGC,GaGogny}, as well 
as with an  interaction derived from   the Argonne V18 nucleon-nucleon potential 
with 
the Unitary Correlation Operator Method\;\cite{PR}. In the case of metallic 
clusters, within the jellium model\;\cite{Br}, the modifications found when 
passing from 
RPA to SRPA are very large\;\cite{GC1} and completely spoil the quality of the 
RPA 
results concerning the dipole strength distribution. In this application, 
the limiting case of highly ionized clusters is very striking.  The RPA is 
able to reproduce the exact theoretical prediction on the position of the 
dipole plasmon excitation (at the Mie energy) and the fact that the dipole 
strength is totally concentrated at this energy\;\cite{YCG}. These very nice 
features are 
completely lost in SRPA\;\cite{GC1}. In the nuclear case, one might 
associate such unpleasant result to some inadequacy of the used nuclear 
interaction. This could be especially true for the Skyrme-- and Gogny--based 
SRPA 
calculations because some 
matrix elements appearing in SRPA are not present at the RPA level, for instance 
those 
with three particle and one hole or three hole and one particle indexes.  
Therefore, they do not 
enter at all in the fitting procedure used to fix the parameters of the 
effective interaction. Furthermore, a certain amount of correlations is already 
contained in an effective way in the parameters of these phenomenological 
forces, because the adjustment of the parameters is performed with 
mean--field calculations. When 
correlations are explicitly introduced in models beyond the mean field, 
double--counting problems may arise. Finally, Skyrme and Gogny forces contain 
zero--range terms and this produces in some cases a cutoff--dependence in 
beyond--mean--field models.

However, the problem of the large downwards energy shift exists also in the case 
of metallic clusters\;\cite{GC1} where there is no cutoff dependence and the 
Coulomb interaction does
not contain any parameter\;\cite{Guet}. The origin of the problem must then be 
different.

In deriving the equations of motion in SRPA\;\cite{Ya,Dro} use is made, as in 
RPA, of the quasi--boson approximation (QBA) and it has been 
argued\;\cite{LR,Ta,Ma} that its use in SRPA is even more questionable than in 
RPA. Recently, this was more precisely understood as related to the replacement 
of the correlated ground state with the Hartree--Fock (HF) one, that is 
generated by the use of the QBA. This replacement produces a violation of the 
stability condition at the SRPA level\;\cite{Pa}. 
 A careful analysis of the merits and limits of the SRPA was presented in 
Ref.\;\cite{Pa}. In particular, the violation  of the stability condition
in SRPA\;\cite{Th2} is illustrated and a generalization of the Thouless 
theorem\;\cite{Th} is proven in the case where a correlated ground state is 
used.
Very recently, a subtraction procedure able to remove the double counting 
in beyond--mean--field theories based on effective 
interactions\;\cite{tse2007,tse2013} was applied to the SRPA 
case\;\cite{GAM2015}. 
Reference\;\cite{tse2013} showed explicitly that this subtraction method is also 
able to ensure the stability condition in extended RPA models such as the SRPA. 
In Ref.\;\cite{GAM2015},
 very encouraging results were found for both low--lying and higher--energy 
excitations. In particular, the strong SRPA downward shift is corrected for 
those states having a dominant 1p-1h nature such as the giant resonances, and 
the found results are in much better agreement with the experimental ones, 
compared to the standard SRPA results.

In Ref.\;\cite{GGCS}, an extension of the SRPA including correlations in the 
ground state was tested by applying it to a solvable model and by comparing the 
results with the exact ones. It was also applied to metallic 
clusters\;\cite{GC2}, finding much better results than in SRPA. However, such 
an extended SRPA  is very  demanding from a computational point of view and its 
application to the study of atomic nuclei is hardly feasible. 
  
On the other hand, approaches like the particle--vibration coupling\;\cite{Co}, 
where one uncorrelated particle-hole configuration is coupled to a correlated 
collective mode, do not suffer from the violation of the stability condition 
\cite{Th,Pa} encountered in SRPA. They are however also affected by the cutoff dependence and the double--counting 
problems if phenomenological zero--range interactions are employed.
  
 Extensions of RPA including multiphonon excitations have been applied in 
several contexts. They make use of mapping procedures to replace the fermion 
particle-hole operators by their boson images\;\cite{KM}. The limits of such 
procedure were tested in Ref.\;\cite{GCS} within a solvable model, by comparing 
the exact results with those obtained by truncating the mapping at different 
levels. It was shown that in order to correct for the violations of the Pauli 
principle, higher--order terms are required 
in the boson expansion. This makes not feasible the application of the approach 
to realistic systems. The procedure presented in 
Refs.\;\cite{An,Kn}  allows one to enlarge iteratively the space in which the 
Hamiltonian  is diagonalized by including multiphonon states built by 
Tamm-Dancoff phonon operators remaining in the fermion space.
 
 The approach that we are going to present consists in considering the RPA 
collective fermion operators as building blocks to construct the space of states 
by acting repeatedly on the RPA ground state (that is, the vacuum of the RPA 
operators). The Hamiltonian is then diagonalized in such space. Since the 
fermionic structure of the collective elementary excitations is fully accounted 
for, one may expect to get satisfactory enough results with a basis of states 
containing up to two such elementary excitations.  Thus, one considers, as in 
SRPA, a space containing 1p-1h and 2p-2h configurations, the latter appearing 
only through the two RPA phonon states. Since the 2p-2h sector of the basis 
space is built with those collective excitations which already contain some 
effects of the residual interaction, one might envisage to reduce the dimensions 
of the space by including only some configurations selected, for example, on the 
basis of their  collectivity. When studying the  excitation spectrum one may 
introduce an energy cutoff and consider only those basis vectors whose  RPA 
energy is lower.  This would allow one to reduce the heavy computational effort 
often required within the SRPA. 
 This approach is very much related to that of Refs.\;\cite{An,Kn}. It is 
however to be noted that in such references Tamm Dancoff phonons were used to construct the 
basis states by acting on the uncorrelated HF ground state. We consider here RPA 
operators acting on the correlated RPA ground state. Therefore, even considering 
the mixing of configurations containing up to two phonons,  we take into account 
higher--order excitations. 
We call our model double RPA and use in what follows the acronym DRPA to denote 
it.

The approach is described in the next section. The energy--weighted sum rules 
are shown to be satisfied in Sec. III,  and applications to the nucleus $^{16}$O 
are presented in Sec. IV.  Finally, in Sec. V we draw some conclusions.

\section{The DRPA model}

In RPA one defines the operators
\begin{equation}
 Q^\dagger_\nu = \sum_{ph} (X^\nu_{ph} b^\dagger_{ph} - 
 Y^\nu_{ph} b_{ph})
 \label{q0}
\end{equation}
and $Q_\nu$, such that
\begin{equation}
 Q_\nu  |\Psi_0 (RPA)> = 0
 \label{vacuum}
\end{equation}
and
\begin{equation}
 |\Psi_\nu (RPA)> = Q^\dagger_\nu |\Psi_0 (RPA)>, 
 \end{equation}
where $|\Psi_0(RPA)>$ and $|\Psi_\nu(RPA)>$ are the RPA ground and excited states 
of the system, respectively. The operators $b^\dagger_{ph}$ and $b_{ph}$ create 
and annihilate, respectively, a particle-hole pair ($b^{\dagger}_{ph}=a^\dagger_p a_h$). 
Here and in the following, coupling to total quantum numbers is understood. 
The  $X$ and $Y$ amplitudes appearing in Eq.~(\ref{q0}) are solutions of the equations of motion
\begin{eqnarray} \label{rpa}
   \begin{pmatrix}
     A^{11} & B^{11} \\
     A^{*11} & B^{*11} \\
   \end{pmatrix}
   \begin{pmatrix}
     X^{\nu} \\
     Y^{\nu}  \\
   \end{pmatrix} 
   =
   E_\nu
    \begin{pmatrix}
     G^{11} & 0 \\
     0 & -G^{*11} \\
   \end{pmatrix}
    \begin{pmatrix}
     X^{\nu} \\
     Y^{\nu}  \\
   \end{pmatrix} ,
\end{eqnarray}

where\;\cite{Ro,RS}
\begin{eqnarray}
 A^{11}_{ph,p'h'} & = <\Psi_0 (RPA)|[b_{ph},[H,b^\dagger_{p',h'}]]
 |\Psi_0 (RPA)> \nonumber \\
 &\simeq <HF|[b_{ph},[H,b^\dagger_{p',h'}]]|HF>,
 \end{eqnarray}
\begin{eqnarray}
 B^{11}_{ph,p'h'} & =- <\Psi_0 (RPA)|[b_{ph},[H,b_{p',h'}]]
 |\Psi_0 (RPA)> \nonumber \\
 &\simeq -<HF|[b_{ph},[H,b_{p',h'}]]|HF>,
 \label{b11}
 \end{eqnarray}

and
\begin{eqnarray}
 G^{11}_{ph,p'h'} & = <\Psi_0 (RPA)|[b_{ph},b^\dagger_{p',h'}]
 |\Psi_0 (RPA)> \nonumber \\
 &\simeq <HF|[b_{ph},b^\dagger_{p',h'}]|HF> = \delta_{pp'}\delta_{hh'}.
 \end{eqnarray}

As explicitly shown, in order to calculate the matrix elements appearing in 
Eq.~(\ref{rpa}), use is made of the  QBA, replacing the RPA ground  state 
($|\Psi_0 (RPA)>$) with the HF one ($|HF>$).

 In SRPA, the excitation phonon operators are generalized to
\begin{eqnarray} \label{qstor}
 &\mathcal{Q}^{\dagger}_{\rho} = 
 \sum_{ph} (X^\rho_{ph} b^\dagger_{ph} - Y^\rho_{ph} b_{ph}) +
\nonumber \\
& \sum_{\substack{p_1h_1\\p_2h_2}} (X^\rho_{p_1h_1p_2h_2} b^\dagger_{p_1h_1}b^\dagger_{p_2h_2} - 
 Y^\rho_{p_1h_1p_2h_2} b_{p_2h_2}b_{p_1h_1}),
\end{eqnarray}
which contain 2p-2h components in addition to the 1p-1h ones. 
This extension allows one to go beyond the harmonic approximation of the RPA
and to describe the spreading width of the excited states through the coupling of the 1p-1h
configurations with the 2p-2h ones.
As in RPA, one assumes
\begin{equation}
 |\Psi_\rho (SRPA)> = \mathcal{Q}^\dagger_\rho |\Psi_0 (SRPA)>
 \end{equation}
and
\begin{equation}
 \mathcal{Q}_\rho |\Psi_0 (SRPA)> = 0 \>,
 \end{equation}
where $|\Psi_\rho (SRPA)>$ and $|\Psi_0 (SRPA)>$ are the SRPA excited and ground states, respectively.
The amplitudes appearing in Eq.~(\ref{qstor}) and the energies $E_\rho$ of the excited states are solutions of
\begin{eqnarray} \label{rpastor}
   \begin{pmatrix}
     \mathcal{A} & \mathcal{B} \\
     \mathcal{B^*} & \mathcal{A^*} \\
   \end{pmatrix}
   \begin{pmatrix}
     \mathcal{X^\rho} \\
     \mathcal{Y^\rho}  \\
   \end{pmatrix} 
   =
   E_\rho
    \begin{pmatrix}
     \mathcal{G} & 0 \\
     0 & -\mathcal{G^*} \\
   \end{pmatrix}
    \begin{pmatrix}
     \mathcal{X^\rho} \\
     \mathcal{Y^\rho}  \\
   \end{pmatrix}, 
\end{eqnarray}
where
\begin{equation}
 \mathcal{X}_\rho =
     \begin{pmatrix}
     X^{\rho}_{ph} \\
     X^{\rho}_{p_1h_1p_2h_2}
   \end{pmatrix} \> \> ,
 \mathcal{Y}_\rho =
     \begin{pmatrix}
     Y^{\rho}_{ph} \\
     Y^{\rho}_{p_1h_1p_2h_2}
   \end{pmatrix} 
 \end{equation}
and
\begin{equation}
 \mathcal{A} =
     \begin{pmatrix}
   A^{11}_{ph,p'h'}  & A^{12}_{ph,p'_1h'_1p'_2h'_2}\\
   A^{21}_{p_1h_1p_2h_2,p'h'} & A^{22}_{p_1h_1p_2h_2,p'_1h'_1p'_2h'_2}
   \end{pmatrix}. 
 \end{equation}
Similar expressions may be written for the (super) matrices $\mathcal{B}$ and $\mathcal{G}$. Again, in order to calculate the above matrix elements, one makes use of the QBA, $|\Psi_o (SRPA) > \simeq |HF>$. The sub-matrices $A^{11}, B^{11}$ and $G^{11}$ are the same as the corresponding RPA matrices. The expressions for $A^{12}, A^{21}, A^{22}$ and $G^{22}$ are analogous, except for the presence of 2p-2h operators. Finally, $B^{12} = B^{21}= B^{22} = G^{12} = G^{21} = 0$ 
for density--independent interactions or if one neglects the rearrangement terms generated by the density dependence of the interaction\;\cite{rearra}. 
As said above, complete SRPA calculations (that is, including the coupling of the 1p-1h configurations with the 2p-2h ones and of the 2p-2h among themselves) made for several systems and with different residual interactions give disturbing results that are corrected when the stability condition is ensured\;\cite{GAM2015}. This may be for example guaranteed either by a subtraction method\;\cite{GAM2015} or by using a correlated ground state\;\cite{Pa}, that is, by avoiding the QBA.

In the DRPA, the RPA collective excitations are used as building blocks and we write the excited states as
\begin{equation}
 |\Psi_\alpha> = \big[ \sum_\nu c^\alpha_\nu Q^\dagger_\nu +
\sum_{\nu_1 \leq \nu_2 } d^\alpha_{\nu_1 \nu_2} 
Q^\dagger_{\nu_1} Q^\dagger_{\nu_2} \big] |\Psi_0> 
\equiv \Lambda^\dagger_\alpha |\Psi_0>
 \end{equation}
maintaining fixed the structure of the RPA operators $Q^\dagger_\nu$. If we further make the approximation
\begin{equation}\label{psiapr}
 |\Psi_0> \simeq  |\Psi_0 (RPA)>,
 \end{equation}
then it is true that
\begin{equation}\label{vacu}
 \Lambda_\alpha|\Psi_0> = 0
 \end{equation}
and
\begin{equation}
 <\Psi_\alpha|\Psi_0> = <\Psi_0|\Lambda_\alpha|\Psi_0> = 0, 
 \end{equation}
that is, the ground state $|\Psi_0>$ is orthogonal to the excited states. 
We remark that, once the approximation (\ref{psiapr}) is made, the vacuum condition, Eq. (\ref{vacu}), follows 
immediately from Eq. (\ref{vacuum}),  satisfied by the elementary operators $Q_\nu$ at the RPA level.
By following the same procedure used to derive the RPA equations of motion\;\cite{Ro,RS} one gets 
\begin{equation}
 H|\Psi_\alpha> = E_\alpha|\Psi_\alpha> \, \Rightarrow \nonumber
\label{motion1}
 \end{equation}
\begin{equation}
 [H,\Lambda^\dagger_\alpha]|\Psi_0> = (E_\alpha -E_0)
 \Lambda^\dagger_\alpha|\Psi_0>. \, \nonumber
\label{motion2}
\end{equation}
From  Eq.~(\ref{psiapr}) and from the vacuum condition, Eq. 
(\ref{vacuum}), it follows
\begin{eqnarray} 
\label{eq1}
 & \sum_{\nu} c^\alpha_\nu <\Psi_0|[Q_{\nu'},[H,Q^\dagger_{\nu}]]|\Psi_0>   \\
&+\sum_{\nu_1 \nu_2} d^\alpha_{\nu_1 \nu_2} <\Psi_0|[Q_{\nu'},[H,Q^\dagger_{\nu_1}Q^\dagger_{\nu_2}]]|\Psi_0> \nonumber \\ 
&= (E_\alpha-E_0) c^\alpha_{\nu'} \nonumber
\end{eqnarray}
and
\begin{eqnarray} 
\label{eq2}
 & \sum_{\nu} c^\alpha_\nu <\Psi_0|[Q_{\nu'_2}Q_{\nu'_1},[H,Q^\dagger_{\nu}]]|\Psi_0>  \\
&+\sum_{\nu_1\nu_2} d^\alpha_{\nu_1\nu_2} <\Psi_0|[Q_{\nu'_2}Q_{\nu'_1},[H,Q^\dagger_{\nu_1}Q^\dagger_{\nu_2}]]|\Psi_0> \nonumber   \\
&=(E_\alpha-E_0) \sum_{\nu_1\nu_2}d^\alpha_{\nu_1\nu_2} <\Psi_0|Q_{\nu'_2}Q_{\nu'_1},Q^\dagger_{\nu_1}Q^\dagger_{\nu_2}]|\Psi_0>.\nonumber
\end{eqnarray}
By using Eq.~(\ref{psiapr}) and by employing the QBA at the RPA level, the matrix elements appearing in Eqs.~(\ref{eq1}) and (\ref{eq2})
can be expressed in terms of the $X$ and $Y$ RPA amplitudes and of matrix elements similar to the SRPA matrix elements $A^{ij}$ and $G^{ij}$, with the HF ground state replacing $|\Psi_0(SRPA)>$. 

We can rewrite Eqs. (\ref{eq1}) and (\ref{eq2}) as   
\begin{eqnarray} 
  \sum_{\nu} c^\alpha_\nu A_{\nu' \nu }+   
\sum_{\nu_1 \nu_2} d^\alpha_{\nu_1 \nu_2} A_{\nu',\nu_1 \nu_2} = (E_\alpha-E_0) c^\alpha_{\nu'} 
\label{primasempli}
\end{eqnarray}
and
\begin{eqnarray} 
 & \sum_{\nu} c^\alpha_\nu A_{\nu'_2 \nu'_1,\nu}+ 
\sum_{\nu_1\nu_2} d^\alpha_{\nu_1\nu_2} A_{\nu'_2 \nu'_1,\nu_1 \nu_2} = \\
&(E_\alpha-E_0) \sum_{\nu_1\nu_2}d^\alpha_{\nu_1\nu_2} G_{\nu'_2 \nu'_1,\nu_1 \nu_2},\nonumber
\label{secondasempli}
\end{eqnarray}
with 

\begin{equation} \label{a11}
A_{\nu'\nu} \equiv <HF|[Q_{\nu'},[H,Q^\dagger_{\nu}]]|HF> = 
E^{RPA}_\nu \delta_{\nu\nu'},
\end{equation}
\begin{eqnarray} \label{a12}
&A_{\nu',\nu_1\nu_2} \equiv <HF|[Q_{\nu'},[H,Q^\dagger_{\nu_1}Q^\dagger_{\nu_2}]]|HF>  \nonumber\\
&=\sum_{\substack{p'h'\\p_1h_1p_2h_2}}(X^{\nu'}_{p'h'}X^{\nu_1}_{p_1h_1}X^{\nu_2}_{p_2h_2}-Y^{\nu'}_{p'h'}Y^{\nu_2}_{p_1h_1}Y^{\nu_1}_{p_2h_2}) \nonumber \\ 
&\times A^{12}_{p'h',p_1h_1p_2h_2},  
\end{eqnarray}
\begin{equation} \label{a21}
A^{21}_{\nu'_1\nu'_2,\nu} = A^{12}_{\nu,\nu'_1\nu'_2},
\end{equation}
\begin{eqnarray} \label{a22}
&A^{22}_{\nu'_1\nu'_2,\nu_1\nu_2} \equiv <HF|[Q_{\nu'_2}Q_{\nu'_1},[H,Q^\dagger_{\nu_1}Q^\dagger_{\nu_2}]]|HF>  \nonumber\\
&=\sum_{\substack{p'_1h'_1p'_2h'_2\\p_1h_1p_2h_2}}(X^{\nu'_1}_{p'_1h'_1}X^{\nu'_2}_{p'_2h'_2}X^{\nu_1}_{p_1h_1}X^{\nu_2}_{p_2h_2} \\
&+Y^{\nu'_1}_{p_1h_1}Y^{\nu'_2}_{p_2h_2}Y^{\nu_1}_{p'_1h'_1}Y^{\nu_2}_{p'_2h'_2})
 A^{22}_{p'_1h'_1p'_2h'_2,p_1h_1p_2h_2} + \mathcal {O}(B^{11}), \nonumber
\end{eqnarray}
\begin{eqnarray} \label{g22}
&G^{22}_{\nu'_1\nu'_2,\nu_1\nu_2} \equiv <HF|[Q_{\nu'_2}Q_{\nu'_1},Q^\dagger_{\nu_1}Q^\dagger_{\nu_2}]|HF>  \nonumber \\
&=\sum_{\substack{p'_1h'_1p'_2h'_2\\p_1h_1p_2h_2}}(X^{\nu'_1}_{p'_1h'_1}X^{\nu'_2}_{p'_2h'_2}X^{\nu_1}_{p_1h_1}X^{\nu_2}_{p_2h_2} \\
&-Y^{\nu'_1}_{p'_1h'_1}Y^{\nu'_2}_{p'_2h'_2}Y^{\nu_1}_{p_1h_1}Y^{\nu_2}_{p_2h_2})  
 G^{22}_{p'_1h'_1p'_2h'_2,p_1h_1p_2h_2}. \nonumber
\end{eqnarray}

The matrices in the right--hand sides of Eqs. (\ref{a12})-(\ref{g22}) are denoted and have the same expressions as the 
SRPA $A^{ij}$ and $G^{22}$ matrices, where the HF ground state is used instead of $|\Psi_0 (SRPA)>$.

In Eq. (\ref{a22}), $O(B^{11})$ denotes several terms containing the RPA matrix $B^{11}$
of Eq. (\ref{b11}). 
In the calculations that we present in this work we neglect for simplicity the terms $\mathcal {O}(B^{11})$ 
appearing in Eq. (\ref{a22}). We have checked within an exactly solvable 3--level Lipkin model\;\cite{lipkin} that their contribution to the DRPA energies is negligible. 

\section{Energy Weighted Sum Rules}

 It is very well known that, if $|\Psi_0>$ and $|\Psi_\nu>$ are the exact eigenstates of a system with Hamiltonian $H$, the following equality holds\;\cite{RS}

\begin{equation}
 \sum _\nu E_\nu \mid <\Psi_\nu \mid T\mid\Psi_0 >\mid^2=\frac{1}{2} <\Psi_0 |\big[T,\big[H,T \big] \big] |\Psi_0 >
 \label{Eq:Thou}
\end{equation}
with $T$ any transition operator. The above equality denotes the EWSR. The Thouless theorem \cite{Th}  states that in RPA the EWSR are valid if the sum in the l.h.s. is calculated with the RPA energies and transition probabilities,whereas 
the uncorrelated ground state $| HF >$ is used in the r.h.s.. It was shown  that the EWSR are valid also in SRPA\;\cite{Ya,AL,Pa} . Following step by step the demonstration of Ref. \cite{Ya}, one easily shows their validity also in the present approach.
 For any self--adjoint transition operator $T$ connecting the ground state to the excited ones we can write

 \begin{equation}
  T= \sum_ \alpha <\Psi_\alpha \mid T\mid\Psi_0 > \Lambda_\alpha^\dagger+  h.c.
 \end{equation}

and 
\begin{eqnarray}
 &<\Psi_\alpha \mid T \mid \Psi_0 >=<\Psi_0 \mid\big[\Lambda_\alpha, T\big]\mid\Psi_0 >\nonumber
\\
  &=\sum_\nu c^\alpha_\nu <\Psi_0 \mid\big[Q_\nu, T\big]\mid\Psi_0 > \\
  &+
  \sum_ {\nu, \nu'} d^\alpha_{\nu \nu'} <\Psi_0 \mid\big[Q_\nu Q_{\nu'}, T\big]\mid\Psi_0 >. \nonumber
\end{eqnarray}

 In the spirit of the Thouless theorem the r.h.s. of Eq. (\ref{Eq:Thou}) is approximated as
 \begin{equation}
  <\Psi_0 |\big[T,\big[H,T \big] \big] |\Psi_0 > \approx <HF |\big[T,\big[H,T \big] \big] |HF >.
 \end{equation}

 Since, as can be easily verified

 \begin{equation}
 <HF |\big[\Lambda^\dagger_\alpha,\big[H,\Lambda^\dagger_\alpha \big] \big] |HF >=0, 
\end{equation}

one gets

\begin{eqnarray}
 &\frac{1}{2}<HF |\big[T,\big[H,T \big] \big] |HF >\nonumber \\
 &=\sum_{\alpha \alpha'}
 T_{0 \alpha}T_{0\alpha'}^*
  <HF |\big[\Lambda_\alpha',\big[H,\Lambda^\dagger_\alpha \big] \big] |HF >
  \nonumber \\
 &=\sum_{\alpha \alpha'}
 T_{0 \alpha}T_{0\alpha'}^* \big[ \sum_{\nu \nu'}c_{\nu'}^{\alpha'}c_{\nu}^{\alpha}A_{\nu'\nu}^{11}
 \\
 &+\sum_{\nu' \nu_1 \nu_2}c_{\nu'}^{\alpha'}d_{\nu_1 \nu_2}^{\alpha}A_{\nu',\nu_1\nu_2}^{12}
  +\sum_{\nu \nu_1' \nu_2'}c_{\nu}^{\alpha}d_{\nu_1' \nu_2'}^{\alpha'}A_{\nu_1'\nu_2',\nu}^{12}
\nonumber \\
 &+\sum_{\nu_1 \nu_2 \nu_1' \nu_2'}d_{\nu_1 \nu_2}^{\alpha}d_{\nu_1' \nu_2'}^{\alpha'}
 A_{\nu_1'\nu_2',\nu_1\nu_2}^{22}\big]. \nonumber
\end{eqnarray} 

By acting with $\sum_{\alpha'} T_{0 \alpha'} \sum_{\nu}c^{\alpha'}_{\nu} \sum_{\alpha}T_{0 \alpha} $ on the l.h.s. and r.h.s. of Eq. (\ref{eq1}), and 
with $\sum_{\alpha'} T_{0 \alpha'} \sum_{\nu_{1}^{'} \nu_{2}^{'} }d^{\alpha'}_{\nu_{1}^{'} \nu_{2}^{'}} \sum_{\alpha}T_{0 \alpha} $ 
on Eq. (\ref{eq2}), and summing side by side the obtained equations we get the EWSR

\begin{equation}
 \frac{1}{2}<HF |\big[T,\big[H,T\big] \big] |HF >= \sum_{\alpha}E_{\alpha} |T_{0 \alpha}|^2.
\end{equation}

We also note that, since the transition amplitudes are evaluated, as in RPA, with reference to the HF ground state, one has
\begin{equation}
 T_{0 \alpha}\approx<HF |\big[\Lambda_\alpha,T \big] |HF >=
 \sum_{\nu}c^{\alpha}_{\nu}\sum_{ph}(X^{\nu}_{ph}+Y^{\nu}_{ph})(p|T|h)
\end{equation} 
 In deriving the last equation it has been assumed that $T$ is a one-body operator.

\section{Results for $^{16}$O}

\begin{figure}
\includegraphics[scale=0.35]{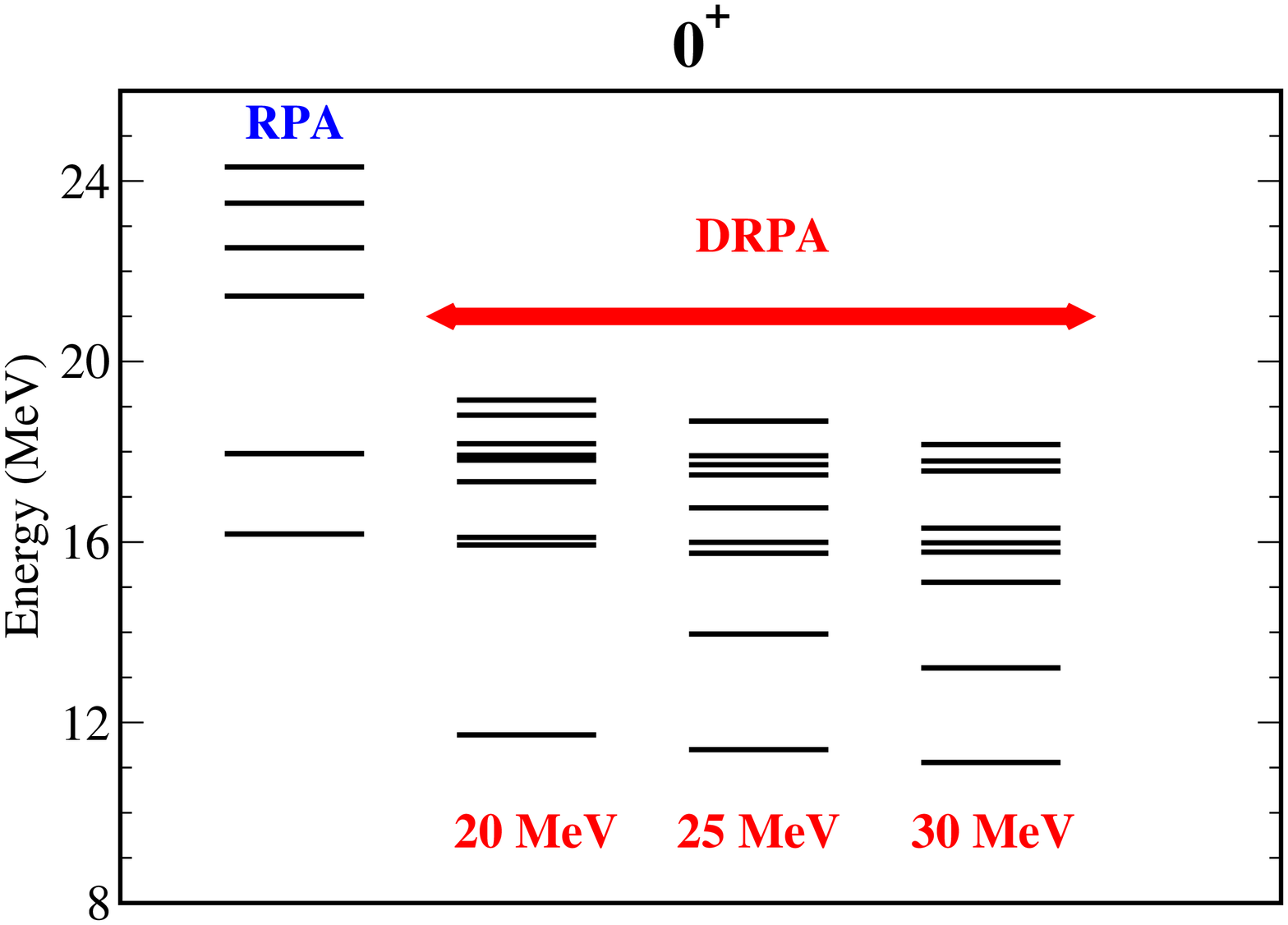}
\caption{(Color online) Monopole states 
calculated with the RPA and the DRPA, with three different values of the cutoff 
 $E1C$, from 20 to 30 MeV and a value of 40 MeV for the $E2C$ cutoff.  }
\label{fig:O16-J0-E1C}
\end{figure}

\begin{figure}
\includegraphics[scale=0.35]{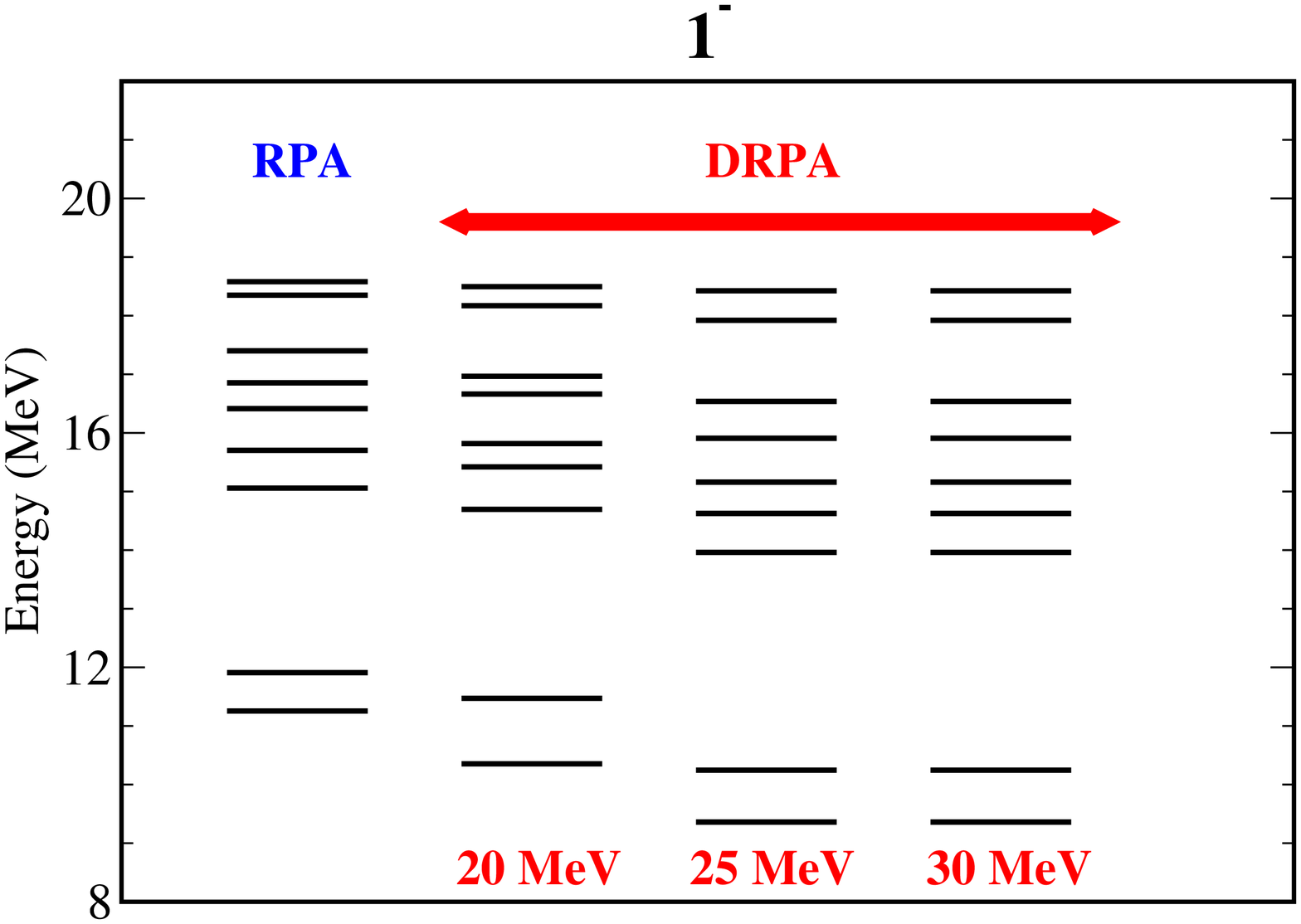}
\caption{ (Color online) Same as in Fig. 1, but for the dipole states. }
\label{fig:O16-J1-E1C}
\end{figure}

\begin{figure}
\includegraphics[scale=0.35]{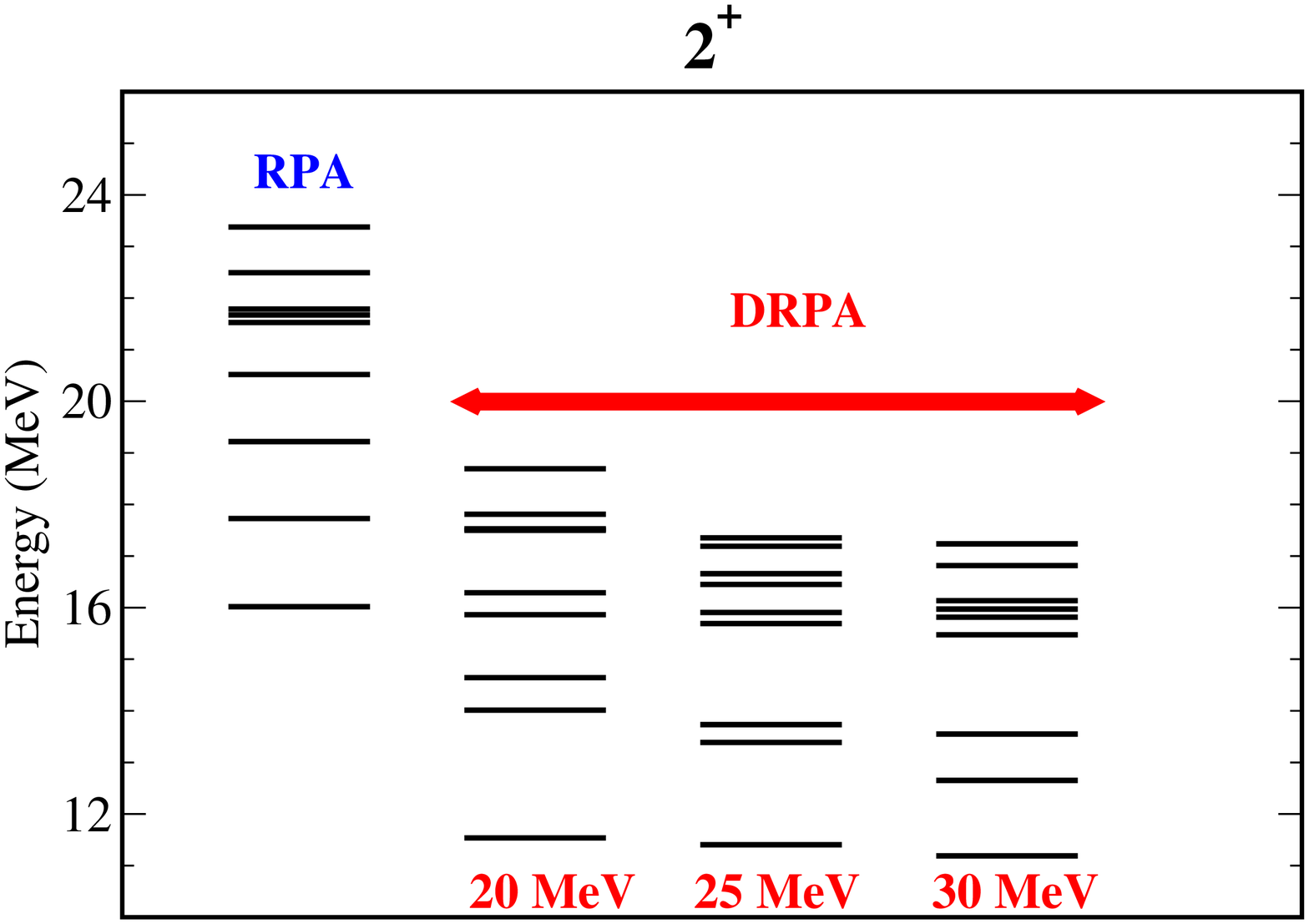}
\caption{(Color online) Same as in Fig. 1, but for the quadrupole states. }
\label{fig:O16-J2-E1C}
\end{figure}

\begin{figure}
\includegraphics[scale=0.35]{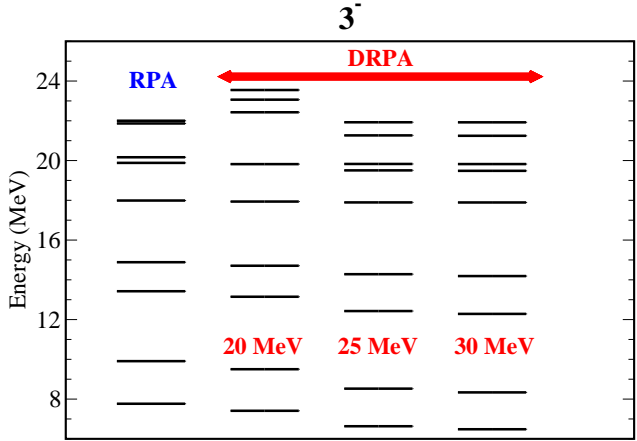}
\caption{(Color online) Same as in Fig. 1, but for the octupole states.}
\label{fig:O16-J3-E1C}
\end{figure}

\begin{figure}
\includegraphics[scale=0.35]{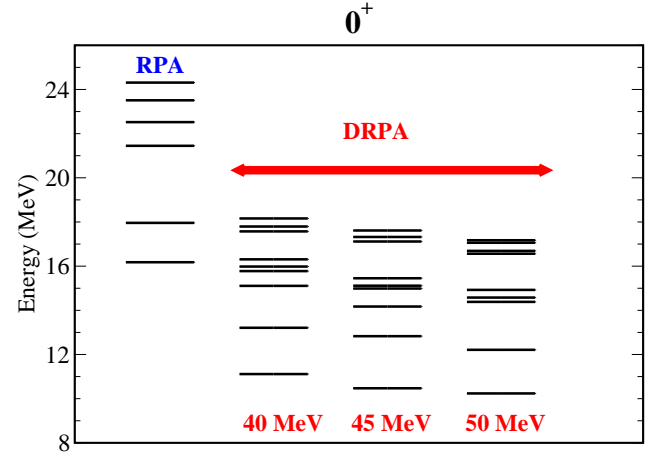}
\caption{(Color online) Monopole states calculated with the RPA and the DRPA, with three different values of the cutoff $E2C$, from 
40 to 50 MeV and a value of 30 MeV for the $E1C$ cutoff. }
\label{fig:O16-J0-E2C}
\end{figure}

\begin{figure}
\includegraphics[scale=0.35]{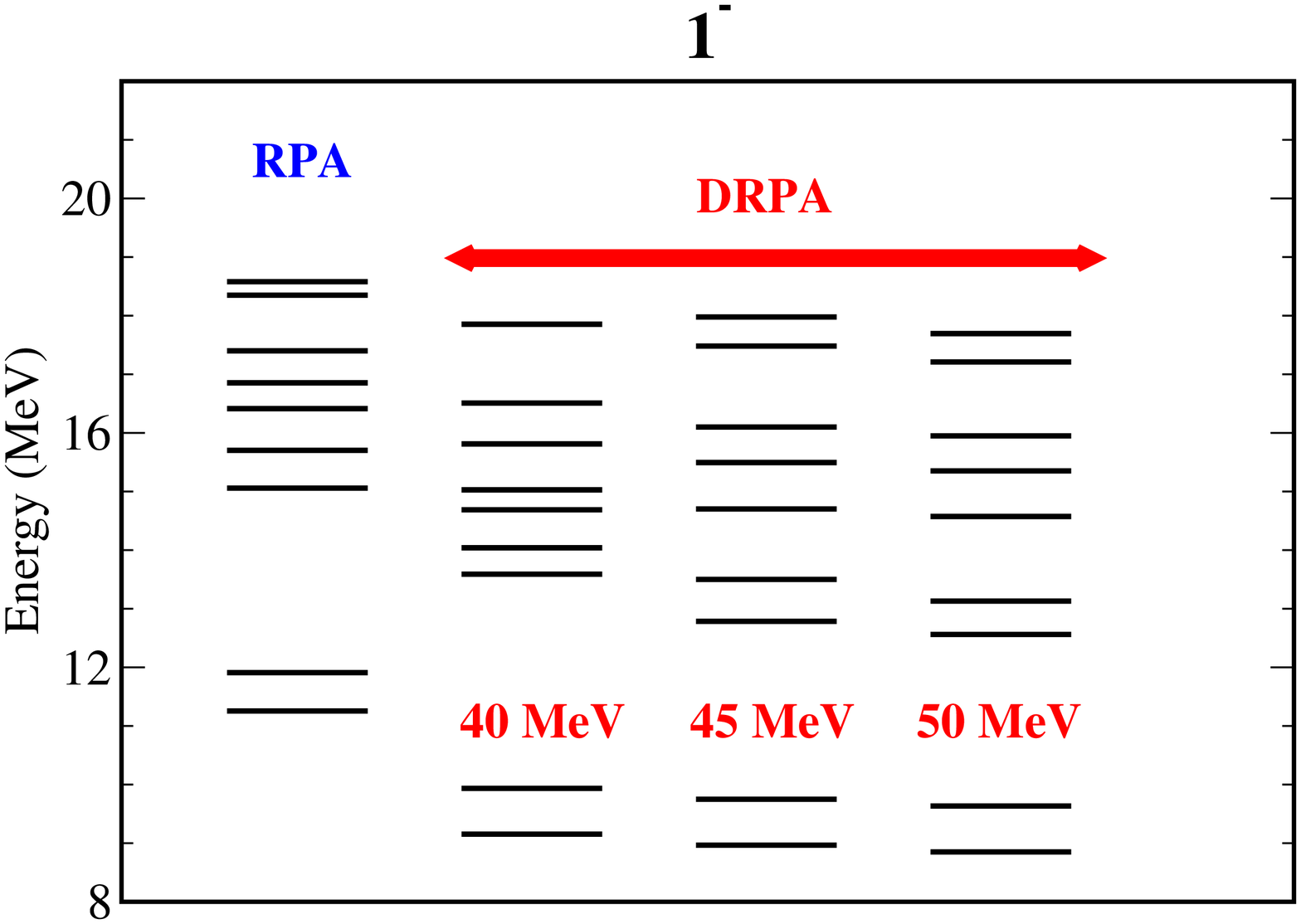}
\caption{(Color online) Same as in Fig. 5, but for the dipole states. }
\label{fig:O16-J1-E2C}
\end{figure}

\begin{figure}
\includegraphics[scale=0.35]{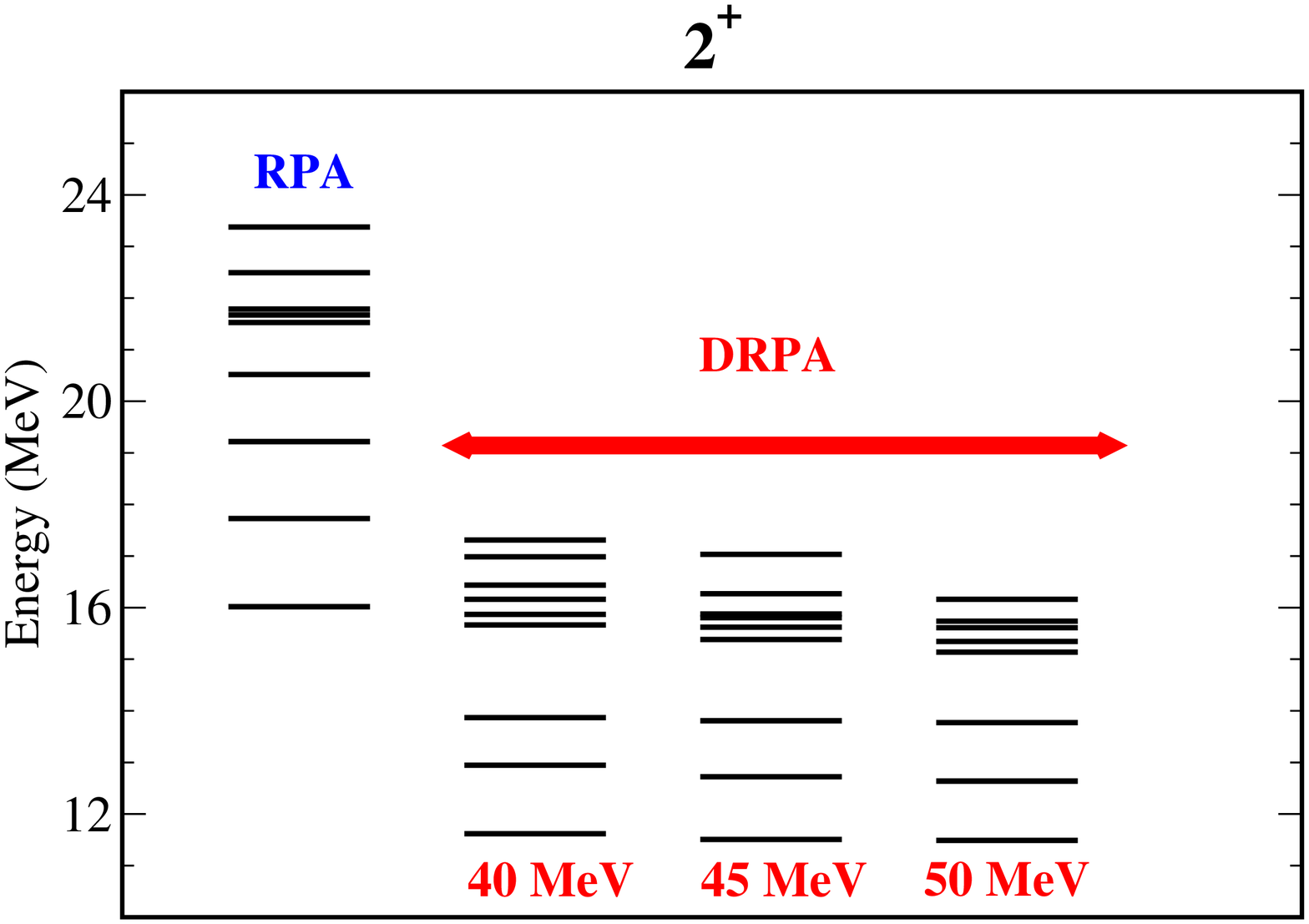}
\caption{(Color online) Same as in Fig. 5, but for the quadrupole states.}
\label{fig:O16-J2-E2C}
\end{figure}

\begin{figure}
\includegraphics[scale=0.35]{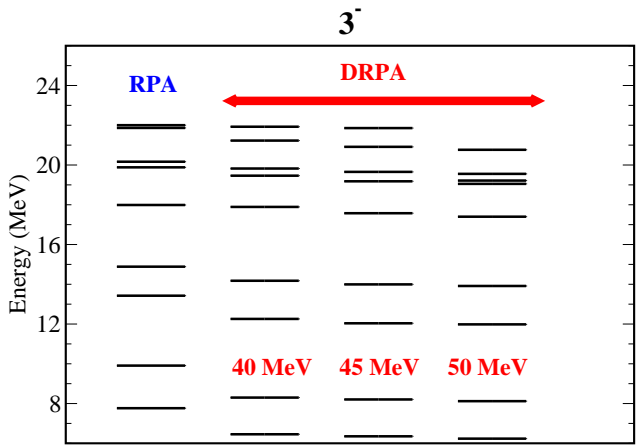}
\caption{(Color online) Same as in Fig. 5, but for the octupole states.}
\label{fig:O16-J3-E2C}
\end{figure}

We have applied the DRPA method to the study of the excitation 
spectrum in $^{16}$O. In this section we present the results of these calculations and
compare them with those obtained in Ref. \cite{GGC} within the SRPA. The latter were shown to be
very different from the RPA ones, also in the parts of the spectrum
which are expected to be reasonably well described by RPA. As in Ref. \cite{GGC}, all
calculations have been done by using the SGII Skyrme interaction\;\cite{sgii}. Starting
with a HF calculation, natural parity RPA elementary excitations are obtained
and used as building blocks to construct a basis of one-- and two--phonon states
with RPA energies up to some given cutoffs, $E1C$ and $E2C$, 
respectively. In order to avoid any misunderstanding due to the use of the word
'phonon', we stress again that we do not make any bosonic mapping of the RPA 
elementary excitations, that is, the fermionic internal structure is maintained and the
Pauli principle is fully preserved. 
First of all, it is necessary to assess the stability of the results with respect
to the values of the cutoffs $E1C$ and $E2C$. In Figs. 
\ref{fig:O16-J0-E1C}--\ref{fig:O16-J3-E1C} we show the
energies obtained with a fixed value of $E2C$ = 40 MeV, with values of $E1C$ running
from 20 to 30 MeV, for the multipole states 0$^+$, 1$^-$, 2$^+$ and 3$^-$. We report also 
the RPA energies. In all cases, as expected, we see that the
energies obtained by diagonalizing the Hamiltonian in the space of one and two
RPA phonons are lower than the RPA results. In the cases of positive parity
states this lowering is very pronounced. This happens because,
for the nucleus we are dealing with, 1p-1h 
configurations of positive parity correspond to a two--major--shell jump and
the corresponding unperturbed energies are quite large. 
The lower--energy states found with the DRPA in the 
0$^+$ and 2$^+$ cases have mostly a 2p-2h 
nature, as was shown with genuine SRPA calculations\;\cite{GAM2015} and cannot thus be found with the RPA. They can be 
predicted here by coupling low--lying negative parity RPA phonons. However, for the
moment, we want to stress that for all multipolarities the results stabilize
with increasing $E1C$ from 20 to 30 MeV. To complete this analysis on the stability
of our calculations, in Figs. \ref{fig:O16-J0-E2C}--\ref{fig:O16-J3-E2C} we show the results obtained by fixing $E1C$ to
the value of 30 MeV and varying $E2C$ from 40 to 50 MeV. Again, the results are 
remarkably stable.  

We stress that, due to the Pauli principle, the set of two phonon states is 
redundant. This means that the associated generalized eigenvalue problem may 
present unphysical solutions if the norm matrix presents singularities. 
The solution of the eigenvalue problem is achieved by using a $QZ$ algorithm
\cite{QZ}. The employed algorithm provides the physical eigenvalue as the ratio of 
two quantities $\alpha$ and $\beta$, the latter 
approaching zero, in the case of singularities of the norm matrix. Analyzing
the $\beta$ quantities, spurious states associated to the redundancy can be identified and 
isolated.
Moreover, we notice that, in general, two different groups of eigenvalues are 
found. A first class, for which the corresponding $\beta$ value is quite stable 
with respect to the increasing of the model space (that is, the single--particle 
basis, and the E1C and E2C cutoff values), and a second one exhibiting instead a 
quite strong dependence on the model space. Such states have beta values that 
are typically much smaller than those of the states belonging to the first 
group. In practice, for each solution,  by increasing the model
space dimension (in particular with respect to the $E2C$ cutoff), we studied carefully
the behavior of the $\beta$ values and only those eigenvalues characterized
by stable values have been retained. 

\begin{figure}
\includegraphics[scale=0.35]{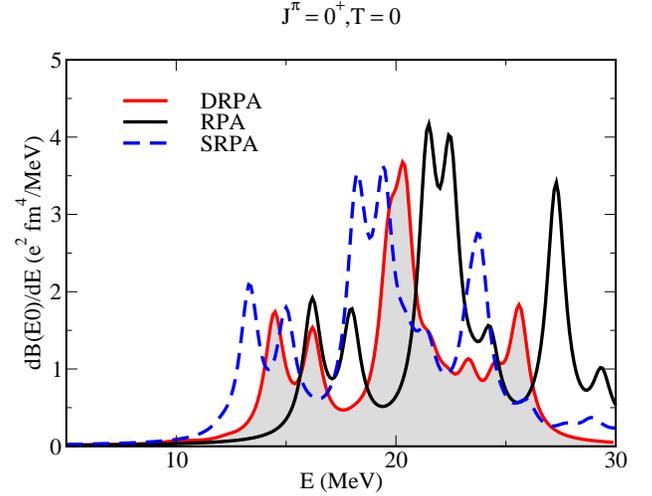}
\caption{(Color online) Monopole response obtained with the RPA (black solid line), the SRPA 
(blue dashed line), and the DRPA (red solid line and grey area). }
\label{fig:O16-J0-response}
\end{figure}

\begin{figure}
\includegraphics[scale=0.35]{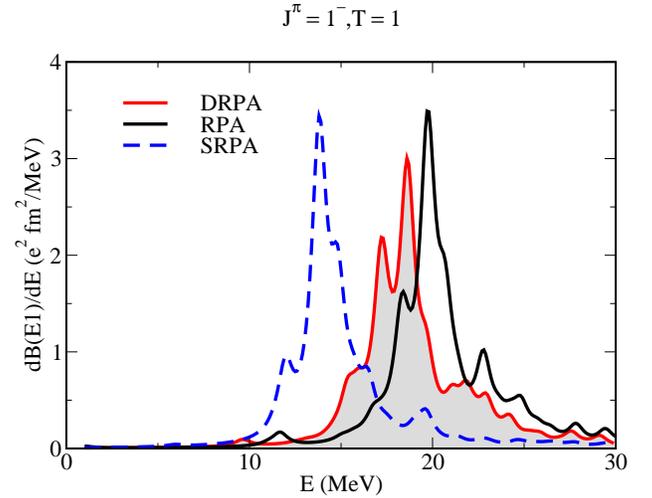}
\caption{(Color online) Same as in Fig. 9, but for the dipole response.}
\label{fig:O16-J1-response}
\end{figure}

\begin{figure}
\includegraphics[scale=0.35]{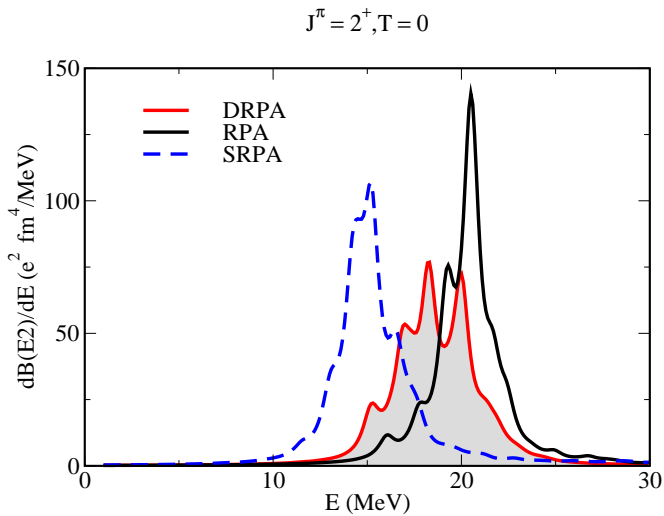}
\caption{(Color online) Same as in Fig. 9, but for the isoscalar quadrupole response. }
\label{fig:O16-J2-response}
\end{figure}

\begin{figure}
\includegraphics[scale=0.35]{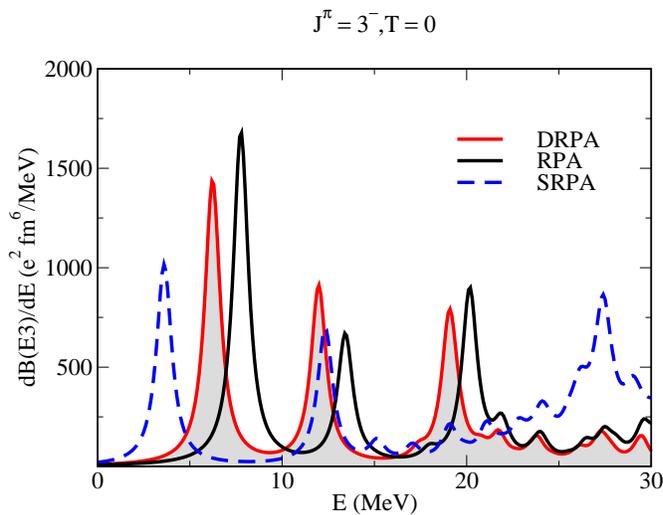}
\caption{(Color online) Same as in Fig. 9, but for the isoscalar octupole response. }
\label{fig:O16-J3-response}
\end{figure}

Let us now turn our attention to the multipole strength distributions. The 
results within the present approach, with $E1C =$ 30 MeV and $E2C =$ 50 MeV are 
presented in Figs. \ref{fig:O16-J0-response}--\ref{fig:O16-J3-response}. 
The shown curves have been obtained by folding the
discrete spectrum with a Lorentzian having a 1 MeV width.
In the same figures the RPA and SRPA results are shown.

Due to the one--body nature of the transition operators, the states having 
a vanishingly small one--phonon component cannot be excited from the ground 
state. This is the reason why almost no strength is seen below 
$\sim$12 MeV in the DRPA response for the $0^+$ and $2^+$ cases. 
As a general
trend, we see in all cases a shift of the DRPA strength distribution
with respect to the RPA, but much less pronounced than in the SRPA. This is a quite
satisfactory result because the positions of the main peaks of the isovector and 
isoscalar
GR remain not much lower than in the RPA, 
and it is known that these excitations are in general well described 
by the RPA model. 
Our present results present a shift, with respect to the RPA energies, that is 
similar to that found
with the subtracted SRPA scheme of Ref. \cite{GAM2015}, that provided a 
satisfactory agreement with the corresponding experimental responses \cite{Young}.

In particular,  
the lowest $3^-$ 
level is predicted at an energy very close to the experimental value \cite{expdata}. In Fig. 12
we see that the SRPA response shows a wide peak centered around 27 Mev, which 
is not present in RPA and DRPA. This comes from some RPA high-lying energy strength
that is shifted downward in SRPA.

\begin{figure}
\includegraphics[scale=0.35]{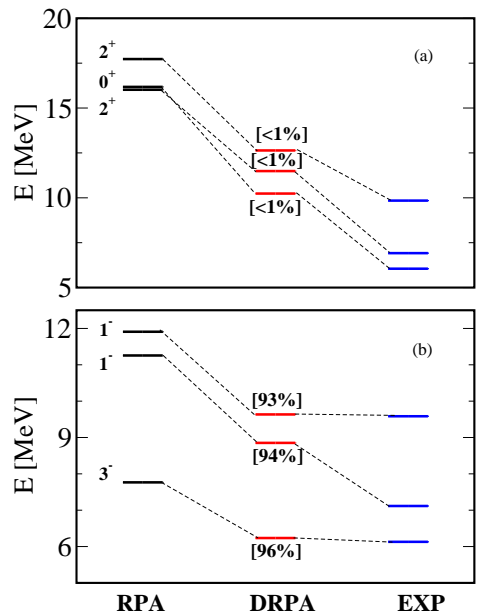}
\caption{(Color online) Comparison of the positive (upper panel) and negative (lower panel) parity low-lying states obtained in RPA and DRPA with the experimental ones. In the DRPA
case, the values in brackets indicate the corresponding one--phonon percentage in the wave functions.}
\label{fig:TheoryVsExp}
\end{figure}

In Fig.\;\ref{fig:TheoryVsExp} we show some positive (upper panel) and negative (lower panel) parity low--lying levels obtained in DRPA and the corresponding experimental ones 
\cite{expdata}.
The numbers in brackets indicate their one--phonon  content (and equivalently the corresponding 1p1h one). The RPA energies are also reported. For the negative parity states the agreement of the DRPA positions with the experimental values is fairly good and the corrections from RPA quite important. On the contrary, for the positive parity states despite the large lowering from the RPA energies, the DRPA values remain still quite different from the experimental ones. This is an indication of the need to include explicitly at least 2p-2h configurations 
besides those already taken into account within the DRPA through the action of one and two RPA phonons on the ground state.

\section{Conclusions}

We have presented an extension of the RPA model, denoted here as DRPA, allowing us to include 
approximately 2p--2h configurations, in a less explicit way compared to the SRPA model. 
The DRPA uses RPA phonons (up to two) as building blocks to construct the excited states. 

We have presented an application to low--lying states and giant resonances in the nucleus 
$^{16}$O. 

In the case of giant resonances, the comparison with the RPA and the SRPA results 
indicates that (i) there is a global downwards shift of the energies with respect to the RPA spectra; (ii) this shift is 
however strongly reduced with respect to that produced by the standard SRPA.  
 The DRPA results are in reasonably good agreement with the experimental
responses. 

This method represents 
for these cases an alternative way to include multiparticle--multihole configurations 
and to correct, at the same time, for 
the anomalous 
shift generated by the SRPA. 
Also for the low--lying states there is a global downwards shift with respect to the RPA results. 
The 
1$^-$ and 
3$^-$ states are in good agreement with the experimental values. The 0$^+$ and 2$^+$ 
states, however, have larger energies with respect to the measured 
values, even if the correction with respect to the RPA goes in the good direction, 
lowering the energies. This discrepancy with the experimental values may be 
understood as due to the almost pure 2p--2h nature of the low--lying 0$^+$ and 2$^+$ 
states. The implicit inclusion of 2p--2h configurations induced by the DRPA is not 
enough to well describe these excitations, and higher--order calculations would be 
necessary.

\begin{acknowledgments}
M.V.A. acknowledges the financial support provided by the
Ministerio de Economía y Competitividad (FIS2013-41994-
P, FIS2011-28738-c02-01), and by Consejería Economía,
Innovación, Ciencia y Empleo, Junta de Andalucía (FQM-160,
P11-FQM-7632).
\end{acknowledgments}

\end{document}